\documentclass[%
reprint,
 amsmath,amssymb,
 aps,
]{revtex4-1}

\usepackage{graphicx}
\usepackage{dcolumn}
\usepackage{bm}

\usepackage{textcomp}

\def\be{\begin{equation}}
\def\ee{\end{equation}}
\def\ba{\begin{eqnarray}}
\def\ea{\end{eqnarray}}

\usepackage[mathlines]{lineno}

\def\dslash#1{#1\!\!\!/\!\, \, }

\def\eps{\epsilon}

\begin{document}

\title{Dark Energy Black Holes with Intermediate Masses at High Redshifts: an earlier generation of Quasars and observations.}

\author{Anupam Singh}
 
 \affiliation{Department of Physics, L.N. Mittal I.I.T, Jaipur, India.}

\date{\today}

\begin{abstract}

Dark Energy is the largest fraction of the energy density of our Universe - yet it remains one of the enduring enigmas of our times.
Here we show that Dark Energy can be used to solve 2 tantalizing mysteries of the observable universe.
We build on existing models of Dark Energy linked to neutrino masses.
In these models Dark Energy can undergo Phase Transitions and form Black Holes.
Here we look at the implications of the family structure of neutrinos for the phase transitions in dark energy and associated peaks in black hole formation.
It has been previously shown that one of these peaks in Black Hole formation is associated with the observed peak in Quasar formation at redshifts $z \sim 2.5$.
Here, we predict that there will also be an earlier peak in the Dark Energy Black Holes at high redshifts $z \sim 18$. 
These Dark Energy Black Holes formed at  high redshifts are Intermediate Mass Black Holes (IMBHs).
These Dark Energy Black Holes at large redshift can help explain both the EDGES observations and the 
observations of large Supermassive Black Holes (SMBHs) at redshifts of 7 or larger.
Our current work directs us to actively look for these Dark Energy Black Holes at these high redshifts as predicted here through targeted searches 
for these Black Holes at the redshifts $z$ near 18. There is a slight dependence of the location of the peak on the lightest neutrino mass. This may enable a measurement of the lightest neutrino mass - something which has eluded us so far.
Finding these Dark Energy Black Holes of Intermediate Mass should be within the reach of upcoming  observations - particularly with the James Webb Space Telescope  - but perhaps also through the use of other innovative techniques focusing specifically on the redshifts $z$ around 18.

\end{abstract}


\maketitle

\section{INTRODUCTION}


Cosmology and Astrophysics were once data starved sciences, today they have become data driven sciences.
This has pushed Cosmology, Astrophysics and fundamental physics much closer together - benefitting progress in all of these exciting areas.
Thus,  the measurements of the Hubble constant and the age of the universe forced us to
re-visit the simplest cosmological model
- a flat universe with a zero cosmological constant\cite{Pierce,Freedman}.
These observations led us\cite{Singh} to the idea of
a non-vanishing vacuum energy due to light fields as playing a significant role in the Universe.
We now have a substantial accumulation of research that has bolstered our belief in what we now call dark energy. This work brings together cutting edge observational work in astrophysics and cosmology and theoretical work in fundamental areas of physics including field theory, elementary particle physics and gravitation. For an exhaustive and elaborate discussion on Dark Energy and related issues, please see  the review \cite{MarkAlessandra} .
{\bf 
It should be noted that what we now call the standard cosmological model uses a small non-vanishing cosmological constant ($\Lambda$) i.e. a non-zero $\Lambda$ and assumes that the cosmological constant problem can be solved in some way \cite{LambdaCDM}.
}

Studies of Dark Energy clusterings have been undertaken to probe dark energy fluctuations - for a review, please see e.g.\cite{DEClustering}.
Even though there was originally some scepticism about Black Hole formation from Dark Energy, recently since the publication of \cite{DarkEnergyCollapseAndBHs} there has been a growing interest in this topic. Thus, for instance, there is detailed work now that can be used to probe Dark Energy Black Holes - please see e.g.\cite{DEblackholeProbe1,DEblackholeProbe2}.

From the perspective of field theory and elementary particle physics, Dark Energy is a low energy phenomenon and so a description in terms of fields in curved space-time can be used to understand many aspects of Dark Energy including the current observational implications of Dark Energy. Here, we study  Dark Energy and its observational consequences by  taking a field theory approach. Field theory coupled to the Standard Model of Elementary Particle Physics has been extremely successful in helping us understand cosmology upto extermely high energies - please see for example\cite{rockymike}. Indeed, at least the expectation was that it would continue to help us understand the low energy universe we currently inhabit. In this context,  observations of Dark Energy raised some uncomfortable questions.
However, a small extension of the Standard Model may help us understand Dark Energy and its consequences. In the Standard Model of Particle Physics, neutrinos are massless particles. A common extension of the Standard Model is to give the neutrinos a mass. Indeed, massive neutrinos have long been considered to provide the most promising window to the physics beyond the Standard Model. This led to an intense world-wide observational program for the search of neutrino masses resulting in the discovery and measurement of neutrino masses - please see for example\cite{NeutrinoMasses} for a recent review. It turns out, as we will discuss below, that massive  neutrinos may indeed also provide a promising pathway forward to build on top of the standard model of cosmology. In particular, the physics of massive neutrinos may help us understand the phenomena associated with Dark Energy and its observational consequences as discussed below.

Field theory models for dark energy motivated by particle physics were discussedl earlier \cite{Singh}.
These fields involved must have very light masses so as to be cosmologically relevant today or in the recent past.
Investigations of particle physics models with naturally  light masses that can generate interesting cosmological consequences at late times
have been done by several authors\cite{GHHK, HolSing}. 
Thus, it was noted that the natural models in which we can get interesting effects are models of neutrino masses with 
Pseudo Nambu Goldstone Bosons (PNGB's). 
The mass scales in these
models can be related to the neutrino masses. Further, any tuning if required can be protected from radiative corrections by the symmetry underlying the Nambu-Goldstone bosons\cite{'thooft}.

PNGBs have a celebrated history in the annals of cosmology because at the late times relevant for observational cosmology any dynamical fields must have small masses compared to the usual masses encountered in the standard model and PNGBs provide a natural mechanism from obtaining light fields. Of the PNGBs perhaps the most celebrated is the axion (please see e.g. \cite{pww,as,df, qaxion,QaisarAxions,rockymarkanupam} for more details). Of all the particles we currently know and have detected the particles with the smallest non-vanishing masses are the neutrinos and so it should not be surprising that PNGBs linked to neutrino masses can be important for understading the current cosmological dynamics.

We now discuss the neutrino PNGB models which we will be using as the core driver of the ideas in the current work. One of the most compelling reasons to take the neutrino PNGB models seriously is because it provides a solution to one of the most perplexing problems of our time - dark energy. Let us examine the genesis of this issue - it goes back to 2 key observables - the age of the Universe and the Hubble Constant and the incompatibility of these measurements in the absence of dark energy\cite{Pierce,Freedman}.
Further, the relationship between these quantities allows us to measure the current dark energy density of the Universe which at first was a complete mystery but clearly tells us something very important and has a bearing on the discussion that follows.

The relationship between the age of the Universe $t_o $ and the Hubble constant $H_o$ can be expressed as:

\be
t_o = \frac{2}{3} H_o^{-1} \Omega_{DE}^{-1/2} \ln \left[ \frac{1 +
\Omega_{DE}^{1/2}}{(1 - \Omega_{DE})^{1/2}} \right]
\ee

where $\Omega_{DE}$ = $\rho_{DE} / \rho_c$. Here $\rho_{DE}$ is the energy density of Dark Energy and $\rho_c = 3 H_o^2/(8 \pi G)$ is the critical density of the Universe with $G$ being the Newton's constant of gravitation.

Observations then tell us that $\rho_{DE}  \sim  (10^{-47} GeV^4) $.
At one point in time this used to be called the Cosmological Constant problem primarily because the above number is very small compared to most energy scales in particle physics.
A resolution of this was provided in \cite{Singh} based on the following key phenomena:
1) Dark Energy gets reduced over time by the process of dissipation;
2) Dark Energy gets dramatically reduced at every phase transition so that the dark energy at the end of the phase transition is $\sim m^4$ where m is the characterisitic mass/energy scale of the phase transition.

Thus, the value of the dark energy density expected is $\rho_{DE}  \sim m^4$ from the last phase transition. In our case, the energy scale of the last phase transition given the current temperature of the Universe is given by $m_2$ the mass of the second generation of neutrinos. This gives a value of the dark energy density which gets rid of the discrepancy between the measurements of the age of the Universe and the Hubble constant.

 As has previously been discussed in detail in \cite{Singh} the PNGB fields in this model can lead to a rapid expansion of the Universe which is consistent with the observed expansion rate of the Universe. This of course is a requirement for any model of what we now call Dark Energy.

Dark Energy is the largest fraction of the energy density of the Universe. Thus, we expect that Dark Energy can play an important role in multiple facets of the Universe.

Thus, we investigated whether this Dark Energy has other important or interesting effects beyond driving the rapid expansion of the Universe on large scales. Since we have a Dark Energy model arising out of fields with a space-time dependence, we can have a rapid expansion of the Universe on large length scales and at the same time have non-trivial  and interesting space time dependent dynamics on shorter length scales driven by the dynamical length scale of the Dark Energy field as was shown in \cite{DarkEnergyCollapseAndBHs}.

In fact, the physics of these fields is very interesting because it allows for Phase Transitions as the temperature of the Universe drops due to the expansion of the Universe as has previously been shown in \cite{GHHK, HolSing, Quasars}. In particular, Black Holes can be formed in these Phase Transitions as previously discussed\cite{Quasars, DarkEnergyCollapseAndBHs}.

In many of our investigations of Dark Energy, we focus mostly on the gravitational interactions of the Dark Energy. It is the dominant interaction of Dark Energy. In this context, the gravitational waves from Dark Energy constitute a new and interesting signal. Thus, we studied the gravitational waves arising Dark Energy dynamics\cite{DarkEnergyGravitationalWaves}. A result from that paper is useful is that the time period of the gravitational waves emitted is $ \sim 10^5$ years .

In a previous paper we have shown that the gravitational waves emitted by Dark Energy can result in the periodicity of the ice ages through the ellipticity variations of the orbit of the earth\cite{DarkEnergyGravitationalWavesAndIceAges}.

{\bf
At this point, it is worth noting also the recent discovery of the Stochastic Gravitational Wave Background\cite{SGWB} and its possible PNGB explanation together with early galaxy formation\cite{SGWBandPNGB}.
}

Thus, we see that Dark Energy and PNGBs already have some important observational implications and we now seek further observational implications.

So far, we have focused mainly on the observational implications of the second neutrino generation with mass $m_2$ since it's energy scale is the closest to the current energy density of the Universe. 

The work done earlier has shown that the phase transition associated with second neutrino generation with mass $m_2$ results in structure formation and in particular in Quasar Formation with the peak of this structure formation happening at redshifts $z \simeq 2.5$\cite{Quasars}.

We now wish to study the implications of the heaviest neutrino with mass $m_3$. As we will see this will lead us to make a prediction about an earlier phase trasition at a higher temperature when the redshift of the universe was about 18. This actually has some very important consequences for our Universe. As we will see, this earlier phase transition will give us Black Holes with masses,

\begin{equation}
M_{BH} \simeq 4.3 \times 10^4 \times M_\odot\ \times \left( \frac{R}{25pc}\right)^2 
\end{equation} 

where R is the size of the dark energy field configuration that collapses.  It should be noted that the value of $R_c \sim 25pc$ being used here follows from the dynamical length scale for the dark energy fields being considered here as we shall see in detail later.

We now briefly describe each of the following sections which are arranged so that each section builds on the previous sections. The next section (Section 2) will discuss the physics of the PNGBs associated with non-vanishing neutrino masses. The potential for the PNGBs will be explicitly given for any finite temperature. Phase transitions in this model at finite temperature will be discussed. There are three phase transitions and we find that the critical temperature of the phase transitions depends on the neutrino masses $m_1 , m_2, m_3$.
In order to make further progress, we need experimental input on the observed neutrino masses. Thus, Section 3 is devoted to a discussion of neutrino masses and their implications.
By the end of section 3 we are in a position to figure out when the phase transition associated with the third generation of neutrinos having mass $m_3$ will occur. We now turn to what to expect at this phase transition. In Section 4 we address the issue of Black Hole formation in the phase transition discussed in Section 3. In Section 5 we turn to addressing the observational issues associated with the discussions in earlier sections - in particular focusing on BHs and their role as seeds for structure formation in our Universe. We note that the predictions made in our work here are at the threshold of observability in the near future.

\section{Symmetry Breaking due to non-vanishing neutrino masses, Potential for the Pseudo Nambu Goldstone Bosons and Phase Transitions.}

Extensive discusssions of the  dynamics of fields and phase transitions in cosmological settings has been done previously
(see e.g. Kolb and Turner\cite{rockymike} ). Standard discussions also exist for the gravitational dynamics of fields
(see e.g. Weinberg\cite{weinberggr}). We have built on these techniques to study and understand the physics of phase transitions  in a general cosmological setting (please see e.g. \cite{nonequilibrium, cmupitt,  us1,usfrw, chiralpt, chiralpt2})  and applied it to the phase transitions in dark energy and the dynamics of dark energy fields.

Consider now the three light families of neutrinos which we will denote by $\nu_j$ with $j$ being the family index which takes on the values $j = 1,2,3$.
We also introduce the left-handed projections: $\nu_{jL} = (1 - \gamma^5) \nu_j/2 $ and right-handed projection: $\nu_{jR} = (1 + \gamma^5) \nu_j/2 $.

Then, we can write down the Lagrangian for the three light families of neutrinos in the form:

\begin{equation}
{\cal L} = \frac{1}{2}  \partial^\mu \Phi_j \partial_\mu \Phi_j + \bar\nu_j i \dslash \partial \nu_j + m_j \bar\nu_{jL}  \nu_{jR} + g_j \bar\nu_{jL} \Phi_j  \nu_{jR} + \rm{H.c.}
\label{Lagrangian}
\end{equation} 

where repeated indices are summed over and $\Phi_j$ are $U(1)$ complex scalar fields that develop vacuum expectation values (VEVs) given by:
\begin{equation}
\langle \Phi_j \rangle = \frac{f e^{i\phi_j/f}}{\sqrt{2}}
\label{VEVs}
\end{equation} 

This results in a potential for the scalar fields which can be calculated using standard techniques (see e.g.\cite{GHHK,HolSing,Thesis}.)
The zero temeperature part of the potential for the scalar fields can be written down in the form:

\begin{equation}
 V (\phi_1,\phi_2,\phi_3) = - \frac{M_j^4}{16 \pi^2} \ln(M_j^2) + K
\label{ZeroTpotential}
\end{equation} 

where $K$ is a constant and 

\begin{equation}
 M_j^2 = m_j^2 [1+\eps_j^2+2 \eps_j Cos(\phi_j/f) ]
\label{M(phij)}
\end{equation} 

and we have defined $\eps_j$ as $\eps_j = \frac{g_i f}{m_j \sqrt{2}} $.

Further, finite temperature corrections can again be calculated using standard techniques (see e.g.\cite{GHHK,HolSing,Thesis}) to get:

\begin{equation}
\Delta V_T (\phi_1,\phi_2,\phi_3) = - 4\frac{T^4}{2\pi^2} \sum_{j=1}^3 \int_0^\infty dx x^2 \ln\left[1+\exp{-\sqrt{x^2+\frac{M_j^2}{T^2}}} \right]
\label{FiniteTpotential}
\end{equation} 

In the High T approximation this potential can be written down as:

\begin{equation}
\Delta V_T (\phi_1,\phi_2,\phi_3) =  \frac{M_j^4}{16 \pi^2} \ln \left(\frac{M_j^2}{T^2} \right)
\label{HighTpotential}
\end{equation} 

The full potential for the scalar fields is then given by the sum of the zero temperature potential and the finite temperature correction:

\begin{equation}
  V_{Full}(\phi_1,\phi_2,\phi_3) =  V (\phi_1,\phi_2,\phi_3) + \Delta V_T (\phi_1,\phi_2,\phi_3) 
\label{FullPotential}
\end{equation} 

Now the full potential of the scalar fields can be used to determine the presence of phase transitions and the critical temperatures of the phase transitions.
It turns out that there are 3 phase transitions in this model and as might be expected the critical temperatures of these 3 phase transitions $T_1$, $T_2$ and $T_3$ are determined by the 3 masses $m_1$ , $m_2$ and $m_3$.
In particular, we note for use in the following sections that $T_3/T_2 = m_3/m_2$ and that all the 3 phase transitions in this model are second order phase transitions.

Let us now take a closer look at the symmetries and the breaking of the symmetries in this model and also place it in a more general context.

Our model has $U_1(1) \times U_2(1) \times U_3(1)$ symmetry in the limit of the neutrino masses going to zero. Each $U_i$ corresponds to $m_i \to 0$ for each of the family indices $i = 1,2,3$.

When $m_i \neq 0$, the Goldstone Boson $\phi_i$ corresponding to $U_i(1)$ when $m_i \to 0$, picks up a non-trivial potential and becomes a Pseudo Nambu Goldstone Boson (PNGB).

It should be noted that the Standard Model of Particle Physics which has massless neutrinos, actually contains the global $U_1(1) \times U_2(1) \times U_3(1) $ symmetries and these are frequently called "Accidental Flavour Symmetries" because they were not put in by hand but they exist in the Standard Model nonetheless - please see e.g. Cabibi et. al.\cite{SMsymmetries} for further details on this. The $U_1(1) \times U_2(1) \times U_3(1) $ symmetry is of course broken by the introduction of non-zero neutrino masses as we have done above and this leads to the PNGBs as described above.

Of course, we note that Flavour Symmetries and Flavour Symmetry Violation are both extremely important for gaining an insight into fundamental particle physics and in particular the masses and couplings of elementary particles. Thus, for example, Flavour Symmetry helps us understand why the masses and couplings of particles are small compared to the scale of symmetry breaking. In the limit that masses and couplings vanish, the Flavour symmetry is enhanced and thus the Flavour symmetry can be viewed as protecting the small masses and couplings in the face of the large scales associated with symmetry breaking.

The symmetries which we are considering - $U_1(1) \times U_2(1) \times U_3(1) $ - which are also present in Standard Model of particle physics are Global symmetries.
In general, Flavour Symmetries can be either Global or Gauged. If the symmetries are gauged , then one has to carefully keep track of anomalies (see e.g. Berezhiani et.al\cite{Berezhianil}).
Berezhiani et. al. find PNGBs linked to massive neutrinos even in the presence of gauged flavour symmetries\cite{Berezhianil}. In fact, since at low energies we expect the physics and phenomenology to converge to the Standard Model supplemented with massive neutrinos, the expectation is that the model considered by us here will capture the low energy physics and phenomenology we are focused on in the current study. 

Thus, we now turn to study the implications of the breaking of the $U_1(1) \times U_2(1) \times U_3(1) $ as described by us above.

The key takeaway from this section which we will use in later sections is that the critical temperatures $T_2$ and $T_3$ of the phase transitions for the families $i = 2$ and $3$  are related to masses of the neutrinos $m_2$ and $m_3$ by $T_3/T_2 = m_3/m_2$ and that all the phase transitions in this model are second order phase transitions.

To make further progress we need the observational input on the masses of the neutrinos. Thus, we now turn to a discussion on the observed neutrino masses and the implications that follow from that.

\section{The Values of Neutrino Masses and their implications }

A significant amount of progress has been made in determining neutrino masses - particularly using Neutrino Oscillations as a probe of neutrino masses. Please see for example\cite{NeutrinoMasses} for a recent review. From the neutrino oscillation data using the normal hiearchy of neutrino families, we can extract\cite{NeutrinoMasses} the neutrino masses - in limit that the lightest neutrino mass is insignificant comapared to the masses of the heavier neutrinos (i.e. $ m_1 << m_2, m_3 $) - we get  that:

\begin{equation}
m_3 \simeq  5 \times 10^{-2} eV  
\label{m3}
\end{equation} 

and

\begin{equation}
m_2 \simeq 9 \times 10^{-3}  eV . 
\label{m2}
\end{equation} 

We now wish to use this to relate the epochs at which the Phase Transitions occur resulting in the peaks in the associated Black Hole formation. We have already labeled the three families of light neutrinos as 1, 2 and 3 in increasing order of mass, so they will have masses given by   $m_1$, $m_2$ and  $m_3$. Let us now label the critical temperatures associated with the phase transitionsas $T_1$ ,  $T_2$,  $T_3$   and scale factors  associated with the peak in BH formation in these phase transitions  as $R_1$,  $R_2$, $R_3$ respectively. The scale factors can themselves be associated with the corresponding  redshifts $z_1$ , $z_2$, $z_3$  respectively.

Using the physics of the phase transitions and the evolution of temperature, scale fator and redshift in an expanding universe\cite{HolSing,Quasars, rockymike,weinberggr} we note that:

\begin{equation}
\frac{T_3}{T_2} = \frac{m_3}{m_2} 
\label{Temperatures}
\end{equation}

\begin{equation}
\frac{T_3}{T_2} = \frac{R_2}{R_3} 
\label{ScaleFactors}
\end{equation}

and

\begin{equation}
\frac{R_2}{R_3} = \frac{1+z_3}{1+z_2}  . 
\label{Redshifts}
\end{equation} 

From observations of the Quasar distribution with redshift, see for example\cite{QuasarsData}, we get:

\begin{equation}
z_2 \sim 2.5 .
\label{z2}
\end{equation}

The value of $z_3$ has a dependence on the lightest neutrino mass $m_1$ and so the result is displayed as a plot of $z_3$  versus $m_1$ in the figure below.

\begin{figure}[h!]
  \includegraphics[width=\linewidth]{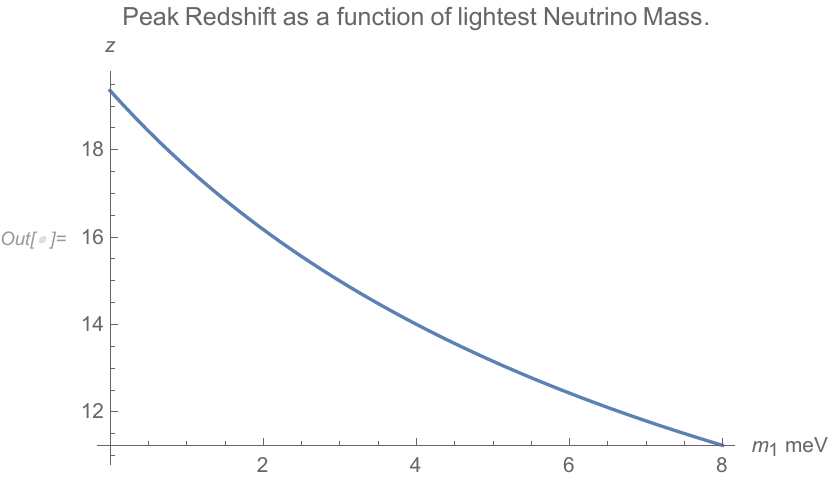}
  \caption{Peak Redshift as a funcion of $m_1$.}
  \label{fig:PeakRedshift}
\end{figure}
Figure \ref{fig:PeakRedshift} displays the loaction of the Peak in redshift (z) as function of the mass of the lightest neutrino ($m_1$ in meV).

If the lightest neutrino mass is insignificant comapared to the masses of the heavier neutrinos (i.e. $ m_1 << m_2, m_3 $), then we get:

\begin{equation}
z_3 \sim 18 . 
\label{z3}
\end{equation} 

It is not possible to extract the value of $m_1$ from neutrino oscillation data, but we can actually evaluate the peak redshift as a function of the mass of lightest neutrino mass this has been done by us. Indeed, the location of the peak at high redshifts may enable us to extract the value of the lightest neutrino mass which has so far eluded us.

At this point it is worthwhile considering the implications of such Black Holes at the relatively high redshifts of $z \sim 18$.
It turns out that this might ease some tensions in existing astronomical data.

\section{Masses of Dark Energy Black Holes formed at High Redshifts}

Before we discuss the data, in order to assess the impact of these Dark Energy Black Holes produced at high redshifts, we also need to arrive at the masses of these Dark Energy Black Holes produced due to the phase transition linked to the mass $m_3$.

Black Hole (BH) formation as a result of cosmological
phase transitions has been etensively studied before by a number of groups.
In general BHs can be formed in both first order and second order phase transitions.
BH formation as a result of first order phase transitions has been studied by R.V. Konoplich et al\cite{Konoplich}, Hawking, Moss, and Stewart\cite{Hawkingetal}, Kodama, Sasaki, and
Sato\cite{Kodamaetal} and Hsu\cite{Hsu}. In First Order Phase Transitions the primary mechanism of BH formation is Bubble Wall dynamics and collisions.
BH formation in second order phase transitions results from the collapse of closed domain walls and this was investigated by Sergei G. Rubin et al\cite{Rubin}  and by Widrow\cite{Widrow}. 
The phase transition we are focused on here is a Second Order Phase Transition and the Black Holes result from the collapse of closed domain walls in this case.
Widrow suggested that the collapse of closed domain walls is likely to produce BHs due to Phase Transitions that occur after the recombination. The mass of such BHs produced has previously been obtained\cite{Quasars,Rubin,DarkEnergyCollapseAndBHs}  and can be stated as:

\begin{equation}
M_{BH} \simeq 4\pi R^2 \times  \sigma .
\label{BlackHoleMassesAsSigma}
\end{equation} 

For the case being considered here $\sigma \sim \alpha m_3^2 f$.
Numerical work studying the gravitational collapse of the Dark Energy fields resulting in the formation of Black Holes\cite{DarkEnergyCollapseAndBHs} gives us the numerical value of $\alpha \simeq 10$.

Black Hole formation in Dark Energy Phase Transitions has been studied earlier \cite{Quasars, DarkEnergyCollapseAndBHs} to arrive at the Black Holes masses. Using the same techniques as earlier but now using the mass of the heaviest neutrino $m_3$  to arrive at the masses of Black Holes formed at the higher redshifts $z_3 \sim 18$, we get:

\begin{equation}
M_{BH} \simeq 4.3 \times 10^4 \times M_\odot\ \times \left( \frac{R}{25pc}\right)^2 
\label{BlackHoleMassesAtHighz}
\end{equation} 

where R is the size of the dark energy field configuration that collapses.  It should be noted that the value of $R_c \sim 25pc$ being used here follows from the dynamical length scale $f/m_3^2$ for the dark energy fields being considered here. Thus, we see that the Dark Energy Black Holes formed at high redshifts come in the desired Intermediate Mass Black Hole category that can get rid of the current tensions in astronomica data.

\section{Meeting the challenges implied by current astronomical data}

In recent years, there has been a recognition that even the existence of quasars at redshifts $z \sim 7$ is at odds with our current theories of supermassive black hole formation - please see for example the recent review\cite{PratoStatementOnSMBHs}.

As noted in this review\cite{PratoStatementOnSMBHs} (and also elsewhere), the observations of quasars and black holes at high redshift and the growth time $t_{growth}$ for black holes for traditional models of black hole formation leads to severe challenges for the traditional models of black hole and structure formation.

To understand the quantitative basis for this it is useful to note the following essential physics involved in the growth of black holes.
The gravitational field of the Black Hole leads to accretion on to the Black Hole but one also needs to consider the radiation pressure of light being emmited by the Black Hole.
Thus, there is a limiting Luminosity called the Eddington Luminosity which is arrived at by balancing the outward force due to radiation pressure by the inward force of gravity.
This equation can be expressed in the form:

\begin{equation}
L_E = K \times M_{BH}
\label{L_eddington}
\end{equation} 
where $L_E$ is the Eddington Luminosity, $M_{BH}$ is the BH Mass and $K$ is a constant.

The total Luminosity from gas accreting onto a BH can be expressed in terms of a radiative efficiency $\epsilon_R$ times the accretion rate:

\begin{equation}
L = \epsilon_R \times \dot{M} c^2
\label{Luminosity}
\end{equation} 
where $\dot{M}$ is the mass accretion rate onto the BH.

The BH accretes the non-radiated component, implying:

\begin{equation}
\dot{M}_{BH} = (1 - \epsilon_R) \dot{M}
\label{Luminosity}
\end{equation}

Thus, inserting the appropriate constants, we get that the growth time $t_{growth}$ for an Eddingtom limited accretion  of a black hole growing from an initial mass $M_0$ to a final mass $M_{BH}$  with radiative efficiency $\epsilon_R$ is given by:

\begin{equation}
t_{growth} \simeq 0.45\frac{\epsilon_R}{1 - \epsilon_R} \ln\frac{M_{BH}}{M_0} Gyr . 
\label{tgrowth}
\end{equation} 

The implication of this is that producing high mass quasars at high redshifts within the available time dictated by the age of the universe becomes a severe challenge for traditional models of black hole formation. This challenge can be overcome if the early phase of black hole formation due to the phase transition linked to neutrinos as described here is taken into account. This is certainly very promising in the context of the ideas discussed in this article. Of course, an explicit observation of such black holes at a redshift $z \sim 18$ would be more compelling and hence searches for black holes at these high redshifts is strongly advocated. This should be within the reach of upcoming observations - particularly with the James Webb telescope - but perhaps also through the use of other innovative techniques  focusing specifically on the redshifts around $z \sim 18$.

It turns out that there is already a hint that such Black Holes may have been formed at the high redshifts described here.
The EDGES instruments have detected an absorption profile in the 21 cm line centred at redshifts spanning approximately $ 15 < z < 20$, centered at $z \sim 17$ \cite{EDGES}.

Dark Energy Black Holes with the Intermediate Masses as described here can easily explain both the EDGES data as described above and also the existence of large SMBHs at redshifts of $z \sim 7$ which can easily result through accretion from the seed Dark Energy IMBHs produced at the redshifts of $z \sim 18$ as described here.

Thus, the Dark Energy Black Holes described here solve 2 of the most perplexing mysteries in the observed universe today. 
At this point, it may be worthwhile to look directly for these IMBHs and our current analysis shows that focusing on redshifts $z \sim 18$ in a search targeted in redshift space may help arrive at an early detection of these Black Holes.

\section{Summary and Conclusion}

A major takeaway from our discussion here is that Dark Energy can provide a source of Black Holes at high redshift. 
Further, we note that the masses of the Dark Energy Black Holes at high redshifts that follow from our calculations place these Black Holes in the desired Intermediate Mass Black Hole category. 
This can help alleviate existing tensions in astronomical data. Further, it provides a target in redshift space to focus on as we look for high redshift black holes.
We have predicted here the redshifts  $z \sim 18$ at which these Intermediate Mass Black Holes should be formed.
Searches for black holes at these high redshifts is strongly advocated. This should be within the reach of upcoming observations - particularly with the James Webb Space Telescope - but perhaps also through the use of other innovative techniques focusing specifically on the redshifts around $z \sim 18$.


\end{document}